# MODELLING RESIDENTIAL SUPPLY TASKS BASED ON DIGITAL ORTHOPHOTOGRAPHY USING MACHINE LEARNING

Klemens SCHUMANN[1,2], Luis Böttcher[2], Philipp Hälsig[1], Daniel Zelenak[2], Andreas Ulbig[1,2]
[1]Fraunhofer Center Digital Energy – Germany
[2]IAEW at RWTH Aachen University – Germany
klemens.schumann@fit.fraunhofer.de

## ABSTRACT

*In order to achieve the climate targets, electrification of individual mobility is essential. However, grid integration of electrical vehicles poses challenges for the electrical distribution network due to high charging power and simultaneity. To investigate these challenges in research studies, the network-referenced supply task needs to be modeled. Previous research work utilizes data that is not always complete or sufficiently granular in space. This is why this paper presents a methodology which allows a holistic determination of residential supply tasks based on orthophotos. To do this, buildings are first identified from orthophotos, then residential building types are classified, and finally the electricity demand of each building is determined. In an exemplary case study, we validate the presented methodology and compare the results with another supply task methodology. The results show that the electricity demand deviates from the results of a reference method by an average 9%. Deviations result mainly from the parameterization of the selected residential building types. Thus, the presented methodology is able to model supply tasks similarly as other methods but more granular.*

## INTRODUCTION

One important aspect of achieving the climate targets is the electrification of the mobility sector. Therefore, Germany has set a target of ten million electric vehicles and one million charging stations by 2030 [1].

Due to the high charging power and the simultaneity of electric vehicle charging, the growth poses challenges for the distribution network infrastructure, which needs to be investigated especially on low and medium voltage levels. To study the impact, network-referenced supply tasks are needed. These consist of geo-referenced electricity demands, load and generation time series as well as charging profiles for the electric vehicles, among others.

While the modelling of synthetic time series (load and generation) and charging profiles is continuously discussed in research, the geo-referencing of electricity demands is still a challenge due to limited data availability. Previous studies such as [2–6] which tackle this topic use socioeconomic data, geospatial data, digital orthophotographs (DOP), or measurement data.

Socioeconomic data contains information on the population and building structure of a region. Among other things, they indicate number of buildings by building size and construction year. For data privacy reasons, the data is aggregated to grids (e.g. $100m\ x\ 100m$) [7]. Supply task modelling based on socioeconomic data uses population and building statistics. Due to the low granularity, conclusions cannot be drawn about the actual location of individual buildings in the distribution network [2].

Geospatial data represents georeferenced shapes of buildings provided with information about the building. Among others, geospatial data is collected on the map service Open Street Maps (OSM). Besides the outlines of buildings, it sometimes is also indicated whether buildings are residential or commercial buildings [8]. Besides OSM, geospatial data can be accessed from cadastral offices (e.g. [9]). Depending on the data set, the data includes house perimeters up to information about building height and roof shape. Methodologies that use geospatial data as data basis make use of available building information [3, 4]. However, data in OSM is uploaded by users non-commercially so that completeness and accuracy cannot be guaranteed. Cadastral data often needs to be licensed.

DOP are georeferenced aerial photographs. Although information cannot be derived directly from these images, buildings or other objects can be identified with the aid of suitable methods [10]. In Germany, DOP are maintained by cadastral offices (e.g. [11]). This ensures high resolution and availability but needs to be licensed in some cases. Other providers offer georeferenced satellite imagery that can be used similarly to DOP (e.g. [12]). [5] presents a methodology to model supply tasks using various data sources. Besides using geospatial data and socio-economic data, the authors also use DOP. However, they apply the DOP in order to identify existing PV-rooftop systems.

Studies that use measurement data-based supply tasks are often conducted in collaboration with distribution system operators (DSOs) [6]. Here, the DSO's data is used. However, these studies are always limited to the network area and data of the DSO.

Since the methods mentioned are not fully suitable to model georeferenced supply tasks, we present a method to determine georeferenced supply tasks based on DOP using image recognition and different machine learning techniques. As training data, DOP and geospatial data is used. The trained models can be applied just using DOP. Adding information from socioeconomic or statistical data, the electricity demand can be determined. Rather than determining the real supply task, the aim is to determine realistic supply tasks. The result of the method can therefore be used for research studies instead of studies for network operators.

In an exemplary case study, we analyze the quality of the trained models. For this purpose, we first train the models on a training region that has a very high quality of open source data. Then, we apply the trained models to a study region. Finally, we discuss different findings, and compare and evaluate the identified energy demands with results of another supply task model [2].

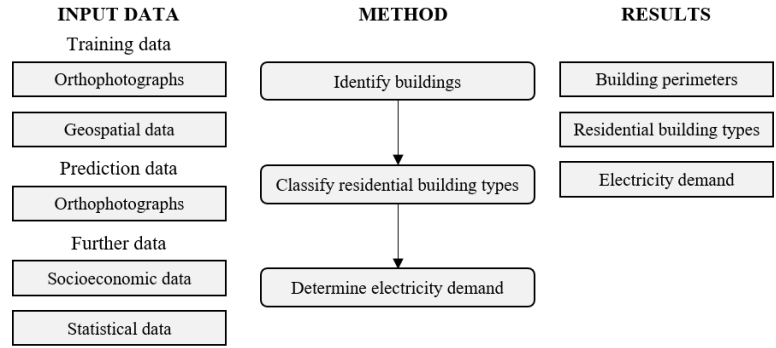

*Figure 1: Overview of the developed methodology*

## METHODOLOGY

### Overview

The aim of the developed method is to determine the electrical energy demand of residential buildings. The prediction using the method should be possible without the use of socio-economic and geospatial data, i.e. it should only require DOP and statistical data. This enables the application in regions for which no reliable socioeconomic or geospatial data is available.

An overview of the procedure is shown in Figure 1, divided into Input data, Method and Results.

The input data is divided into three parts: Training data, prediction data, and further data. The training data is DOP and geospatial data. The prediction data contains only DOP. Socioeconomic data or statistical data is used to determine the electricity demands.

The procedure is split into three steps. In the first step, we identify residential buildings using neural networks with semantic segmentation. In the second step, residential building types are assigned to the buildings. In the last step, we calculate the electricity demand of each building based on the residential building type.

The results are the detected building perimeters, the classification of buildings into residential building types and the electricity demand of each residential building.

### Building identification

The first step aims at identifying residential buildings from DOP. Therefore, DOP and geospatial data is used as training data. The DOP is used as features. From the geospatial data the building floor areas and the assignment to residential are used as labels.

With this input data neural networks with semantic segmentation are trained. We use the framework RasterVision for this purpose [10]. RasterVision allows the automated training and application of neural networks for image recognition.

### Residential building type classification

The goal of the second step is to assign residential building types to the identified residential buildings. Residential building types contain information about the number of households in a building and the floor area of a building, and thus allow estimating the electricity demand for each building [13].

Input data sets for the training are DOP and the identified residential building shapes from step one (Features). Geospatial data containing information on residential building types of each building is the input label for training. Furthermore, socioeconomic data or statistical data can be used to improve the quality of the training data if it is available.

Residential building types are determined based on the size class and construction year of the buildings. Therefore, the determination process first classifies the size classes and second determines the construction year.

The size class represents the type of building, e.g. detached single buildings, row houses, apartment blocks or perimeter block developments. These size classes differ partly by the size and shape of the building shapes. Thus, machine learning methods can be applied to the identified building shapes.

Two different methods are used to classify the size class. The first method uses the RasterVision framework to train neural networks to classify a size class for each building. It uses the building types stored in geospatial data as labels (e.g. in [9]) and DOP as features.

The second method uses a Random Forest classifier trained on building geometry properties to assign the identified building areas to size classes. Again, geospatial data is used as labels to evaluate the results of the Random Forest in training. The features are properties such as the floor area and the convexity of the building, and are determined from the building shapes.

Preliminary examination has shown that the highest quality is achieved when classification is done first with RasterVision. All unassigned buildings are then classified using Random Forests. Using this combination, we could identify the size classes *detached single buildings*, *row houses*, and *perimeter block developments*.

To determine the construction year of each building, information from socioeconomic data or statistical data is used to randomly assign construction years to the buildings. For this purpose, the age distribution given in the available data sets for a region is used (e.g. in [7]).

Combining the size class and the construction year for each building, a residential building type can be selected from [13]. This residential building type contains information about the number of households and the floor area of the reference buildings.

## Electricity demand determination

The aim of the third step is to assign an energy demand to each building using the previously determined information.

Input data for this are the identified buildings with their floor area and the assigned residential building types, the information of the residential building types (e.g. from [13]), and statistical data on the average electricity demand per household.

The electricity demand of each building is then determined using the following formula. $A^{ib}$ is the floor area of the identified building (ib). $A^{rbt}$ is the floor area of the residential building type (rbt). $n_{hh}^{rbt}$ is the number of households of the residential building type and $E_{hh}^{stat}$ is the statistical average electricity consumption of one household. Thus, the average electricity demand is scaled with the number of households and the area of the building.

$$E^{ib} = \frac{A^{ib}}{A^{rbt}} * n_{hh}^{rbt} * E_{hh}^{stat}$$

## CASE STUDY

The presented methodology results in georeferenced buildings and their electrical demand. To check the methods plausibility, first the identification of the buildings and classification of size classes is analyzed. In addition, the results are compared with the results from the method presented in [2].

The DOP training data is obtained from the web map service of the state of North Rhine-Westphalia in Germany [11]. As geospatial data for the building identification we use OSM [8], for the residential building type classification [9]. In terms of socio-economic data, we use [7] for the year of construction and Tabula [13]. The average electricity demand per household is based on [14]. The RasterVision model and the Random Forest are trained on an urban region. The prediction takes place on another region in the same state to obtain the same image quality in the DOP.

## Analysis of the building identification

First, we examine the results of building identification with neural networks. An exemplary section of the study region is shown in Figure 2. The DOP can be seen in the background. The shapes of the identified buildings are marked in blue.

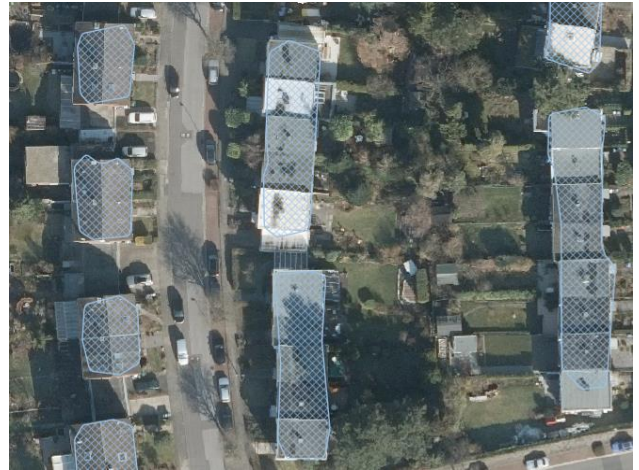

*Figure 2: Exemplary excerpt of the detected buildings*

It is found that all the buildings in the section shown have been identified. In addition, a garage (second uppermost building on the left) and garden sheds (between the middle and right buildings) have not been identified as buildings. Thus, the neural network can distinguish residential buildings from other structures.

However, the model has problems with identification in two places: With the accurate shape of buildings and with contiguous buildings.

From the detached buildings on the left, it can be seen that the buildings are not exactly recognizable as angular. Especially at the corners of the buildings, the neural network tends to represent the building more rounded and to cut off corners. At the connected buildings (middle and right) it can be seen that the model cannot distinguish between single buildings and identifies all connected buildings together as one long building.

The non-exact identification of the building shape should not cause any particular problems in determining the energy demand of a building. The deviation from the total floor area is only small, as can be recognized in Figure 2. The lack of distinction between connected buildings is also not problematic for the determination of energy demands, since the total floor area is also correctly determined in this case.

However, if the identified buildings are to be mapped to low-voltage networks, the lack of subdivision of the buildings could pose a problem since it is difficult to draw conclusions about the number of house connections.

## Analysis of the size class classification

Next, we examine the assignment of residential building types. Here, we focus on the size class classification, since the assignment to year of construction is random. An exemplary section of the size class classification is shown in Figure 3. Here, all buildings were classified into the three classes described in the section methodology. Detached single buildings are shown in green, row houses are shown in orange, and perimeter block developments are shown in pink.

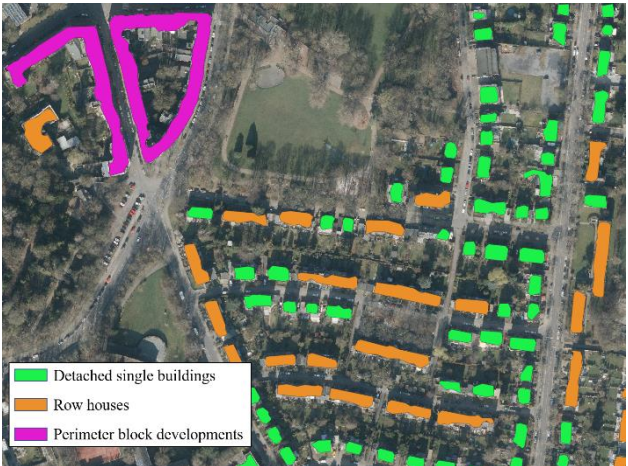

*Figure 3: Exemplary excerpt of the classified size classes*

Distinctions between the three building types are obvious. Detached single buildings are small buildings, row houses are characterized by an elongated shape, and perimeter block developments are elongated buildings that are connected along streets even through corners.

In the case of detached single buildings, however, it is not possible to determine whether the building is a single-family or two-family dwelling. All buildings with a short rectangular structure are classified as detached single buildings. This can make it difficult to identify house connections, as with the other building types.

When classifying the size classes, we found that the classification of the three mentioned size classes works better than others. In particular, apartment buildings or high-rise buildings cannot be well distinguished from the mentioned size classes. This is due to the fact that it is not possible to draw conclusions about the height of the buildings from the DOP. It might be possible to determine the height via the shadow cast, but this would require information about the time when the photo was taken.

**Analysis of the electricity demand**

Finally, the determined electricity demand is to be checked for plausibility. For this purpose, the electricity demand in the study region was determined. As study region we used the area of the city of Aachen, which covers $160\ km^2$. The results from the presented methodology were compared with results from the methodology presented in [2]. The comparison methodology from [2] models the supply task based only on socioeconomic data. Since no georeferencing is available in socioeconomic data, the buildings are assigned only to sub-postal-code areas. This allows modeling supply tasks for system studies, but not modeling of supply tasks for distribution networks (which is focus of this work). Since a comparison with the results from [2] on building level is not possible, the results were compared on postal-code level. For this purpose, the electricity demand of the presented methodology was aggregated. To ensure comparability, the same values were assumed for the annual energy demand of a household.

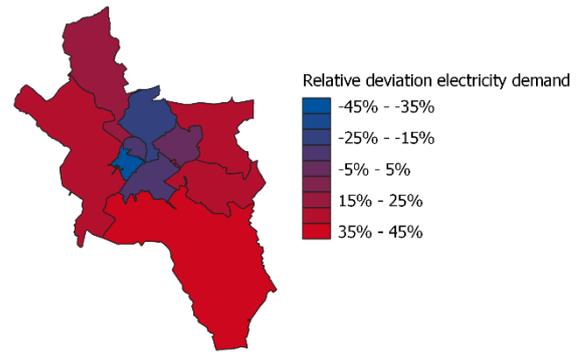

*Figure 4: Relative deviation of aggregated electricity demands in the postal code areas of the study area with [2]. Positive deviation: Demand in presented method lower than in comparison method. Negative deviation: Demand in presented method higher than in comparison method.*

In the study region, 17,500 buildings with an overall floor area of $4.3\ km^2$ have been identified (note that contiguous buildings have been identified as one building). Figure 4 depicts the relative deviation of aggregated electricity demand in all postal-code areas related to the comparison methodology. A positive deviation indicates that the presented methodology has determined a lower electricity demand than the reference methodology and vice versa.

There are deviations between the two results, ranging from $-35\%$ to $42\%$ in the electricity demand related to the reference data set. Over the entire study area, 9% less electricity consumption is determined in our method than in the reference method. This shows that the presented methodology is able to generate similar supply tasks even more granular.

Structural differences can be identified. Using the presented methodology, in the four centrally positioned postal-code areas, the energy demand is estimated to be higher, in the six outer areas lower than in the reference data set. This is due to the choice of residential building types. The centrally positioned regions are inner-city regions. Here, perimeter block developments were predominantly identified. Due to a high ratio of number of households to floor area in the selected residential building types for perimeter block developments, the energy demand in this region is estimated to be higher than in the reference methodology. The outer regions are suburban. These regions are dominated by detached single buildings. In the suburban regions, the energy demand is estimated to be lower due to a low ratio of number of households to floor area compared to the comparison methodology.

In summary, the presented and comparison methodologies provide similar results. The presented methodology offers the advantage that the electricity demand was identified georeferenced.

**DISCUSSION**

In order to study the impact of electromobility on the electric distribution network, information about the residential supply task is needed. In previous research, this supply task is modeled in several ways. However, these are often too aggregated, not complete, or dependent on DSOs as project partners.

In this paper we presented a method that allows the modeling of residential supply tasks based on DOP. This should allow modeling of the supply task even for larger study regions without a good database in geospatial data.

The developed methodology is divided into three steps: building identification, residential building type classification and electricity demand determination. Input data sets are DOP, geospatial data, socio-economic data and statistical data. The result is the supply task for the study region.

The exemplary results have shown that although the electricity demand does not deviate far from the reference data set, the identified buildings are not completely identified, and row houses cannot be subdivided. As a result, no direct conclusions can be drawn about the number and location of house connections. In addition, the developed method can only subdivide between three size classes. In particular, it is not possible to distinguish between row houses and high-rise buildings, since no information on building height is available.

In summary, the method is well suited to determine residential supply tasks. However, it should always be seen as a supplement to existing methods such as [4]. In regions with high quality geospatial data, the results of these methods are usually better.

In future research, we will combine our method with the method presented in [4]. With the combined data sources, a higher quality for supply tasks can be achieved. We will also use the image recognition from DOP for the recognition of other assets, such as PV plants, wind power plants or substations.

## MISCELLANEOUS

### Acknowledgments

This research is part of the project PlaMES (Integrated Planning of Multi-Energy Systems). PlaMES has received funding from the European Union's Horizon 2020 research and innovation programme under grant agreement No. 863922. The content of this publication reflects only the authors' view. The European Climate, Infrastructure and Environment Executive Agency (CINEA) is not responsible for any use that may be made of the information it contains.